\documentclass{article}
\usepackage{array,hhline,amssymb,amsthm}
 
\newcounter{risunok}

\title{Nonassociativity as gravity}
\author{V.~Yu.~Dorofeev\thanks{Friedmann Laboratory For Theoretical Physics,
Department of Mathematics, SPb EF University, Sadovaya 21, 191023
St.Petersburg, Russia, E-mail: friedlab@mail.ru,
dor@vd8186.spb.edu}}
\date{}
\begin{document}
\maketitle
\begin{abstract}
Gravitational interactions are treated as a nonassociative part of
the field interactions. The vector-potential of  Dirac equation is
extended to octonionic algebra. The solution is considered as an
element of the matrix set $BS\times O_4$.
\end{abstract}
\section*{Introduction}
Generalization of physical theories to octonions has its own history
\cite{Fradkin}. Since octonions  form an algebra and have
$U(1)$, $SU(2)$ и $SU(3)$ automorphism groups,  there were attempts
to use this algebra as a basis for the theory of electroweak and strong
interactions.

In this article we propose to consider nonassociativity as gravity
\cite{Dorofeev}. On this way the all-in-one picture of all
interactions arises.

\section{Dirac equation in GRG and on the octonions}
The interaction of  $A_a(x)$ with Dirac particle is introduced as
minimal:
\begin{equation}\label{Dirakm}
(i\gamma^a(\partial_a-iA_a)-m)\Psi(x)=0
\end{equation}
where $\gamma^a$ -- Dirac matrices:
\begin{equation}\label{matDir}
\gamma^0=\left(\matrix{I&0\cr0&-I}\right),\qquad
\gamma^a=\left(\matrix{0&\sigma^a\cr-\sigma^a&0}\right)
\end{equation}
and $\sigma^a,a=1,2,3$ -- Pauli hermitian matrices.

If there is  $SU(2)$  symmetry then:
\begin{equation}\label{Diraksu2}
(i\gamma^a(\partial_a-iA_a^{(b)}\sigma^b)-m)\Psi(x)=0.
\end{equation}

According to Utiyama \cite{Utiyama} Dirac equation with minimal coupling in
curved coordinates leads to the appearance of compensating field
$\Gamma_\mu(x)$ in the following form
\begin{equation}\label{Dirakgr}
p_\mu\to(\partial_\mu-\Gamma_\mu)\Psi(x)=
(\partial_\mu-\omega_\mu^{ab}(x)\sigma^{ab})\Psi(x),
\end{equation}
where $\omega_\mu^{ab}(x)$ -- real functions determined by vierbeins
$e^a_\mu=\partial x^a/\partial u_\mu$ ($x^a$ -- coordinates in flat
space and $u_\mu$ -- in  curved coordinate system), and
$\sigma^{ab}$ -- matrix as $\sigma^{ab}=\frac18[\gamma^a,\gamma^b]$.

The important difference between (\ref{Diraksu2}) and
(\ref{Dirakgr}) is the reality of the coefficient
$\omega_\mu^{ab}(x)$ and formally we "unify" gravity and
electroweak interactions by real $\omega_\mu^{ab}(x)$ and imaginary
$iA_\mu^{a}(x)$. But in fact there is another way to unify all
interactions,  i.e.,  to go beyond the associative algebra.

Really, let the field $A_a(x)=A_a^{\tilde a}(x)\Sigma^{\tilde a}$ is
defined on the nonassociative algebra with generating
$\Sigma^{\tilde a}$. We select octonionic algebra as nonassociative
algebra:
$$\Sigma^0=\left(\matrix{I&0\cr0&I}\right),\quad\Sigma^{\bar
a}=\left(\matrix{0&-\sigma^{\bar a}\cr \sigma^{\bar a}&0}\right),
\quad\bar a=1,2,3,$$
\begin{equation}\label{Z}
\Sigma^4=\left(\matrix{iI&0\cr0&-iI}\right),\quad\Sigma^{4+\bar
a}=\left(\matrix{0&i\sigma^{\bar a}\cr i\sigma^{\bar a}&0}\right),
\quad I=\left(\matrix{1&0\cr0&1}\right),
\end{equation}
and introduce special multiplication "$*$" \cite{Daboul}:
\begin{equation}\label{D}
\left(\matrix{\lambda&A\cr B&\xi}\right)*
\left(\matrix{\lambda'&A'\cr B'&\xi'}\right)=
\left(\matrix{\lambda\lambda'+\frac12tr(AB')\hfill&\lambda A'+\xi'
A+\frac i2[B,B']\hfill\cr\lambda'B+\xi B'-\frac
i2[A,A']&\xi\xi'+\frac12tr(BA')\hfill}\right)
\end{equation}
where $A,A',B,B'$ -- matrices $(2\times2)$ and
$\lambda,\lambda',\xi,\xi'$ -- scalar matrices $(2\times2)$.

Let the vector-potential $A_a^b(x)$ belongs to nonassociative
algebra, i. e. to octonionic algebra. So we come to the matrix
represention:
\begin{equation}\label{Dirako}
(i\gamma^a(\partial_a-iA_a^{(b)}\Sigma^b)-m)\Psi(x)=0.
\end{equation}

Then in case $A_a^4(x)=(0,A(r)\vec r/r)$ and $A_a^i(x)=0,i\ne4$  in
spherical coordinates we have:
\begin{equation}\label{Diraco4}
(\gamma^0\partial_t+\tilde\gamma\partial_r
-\tilde\gamma\vec\Sigma\cdot\hat{\vec L}+im)\Psi(x)=-\tilde\gamma
A(r)\Psi(x),
\end{equation}
where $\vec L=\vec r\times\vec p$ is the angular momentum  operator,
$$\tilde\gamma=\gamma^1\sin\theta\cos\varphi+
\gamma^2\sin\theta\sin\varphi+\gamma^3\cos\theta,\quad
\vec\Sigma=\left(\matrix{\vec\sigma&0\cr0&\vec\sigma}\right).$$

Let's remember this equation and come to Dirac equation on the other
hand.

We introduce curved homogeneous isotropic space. Then the general
form of  static metric looks like:
\begin{equation}\label{ois}
ds^2=K^2(r)dt^2-B^2(r)dr^2-r^2d\Omega^2.
\end{equation}

Particularly, if $K^2=B^{-2}=f^2=1-r_g/r$, $r_g$ -- gravitational
radius -- this is the Schwarzschild metric:
\begin{equation}\label{Shwm}
ds^2=(1-\frac{r_g}r)dt^2-\frac{dr^2}{1-\frac{r_g}r}-
r^2(\sin^2\theta d\varphi^2+d\theta^2).
\end{equation}

In the Dirac equation (\ref{Dirakgr}) we put Ricci rotation
coefficients of  curved space (\ref{Shwm}) \cite{Wheeler}:
\begin{equation}\label{DirakShw}
(\gamma^0\frac1f\partial_t+\tilde\gamma f\partial_r
-\tilde\gamma\vec\Sigma\cdot\hat{\vec L}+im)\Psi=-\tilde\gamma(\sqrt
f(\sqrt f)_{,r}+\frac1r(\sqrt f-1))\Psi.
\end{equation}

Thus we get likeness of Dirac equation in the Schwarzschild metric
and Dirac equation on the nonassociative algebra which in this case
is octonionic algebra. Formally nonassociativity is necessary to
relieve  imaginarity in the long derivative. Indicated likeness
allows us to make supposition: the nonassociativity should be
considered as a local demonstration of the curvature of space.

As we go to  the curved space, we have to introduce vierbein variables:
$$\Delta t\to\Delta\tau=K(r)\Delta t,\quad
\Delta r\to\Delta R=B(r)\Delta r$$ and to give physical meaning to
the new variables.

Remind that  the time dilation and the length contraction in
the special theory of relativity (STR) have the form
$$dt^2=(1-v^2)dt_0^2,\quad dl^2=(1-v^2)dl_0^2.$$

It is natural to suppose: if $r\to\infty$ then functions $K(r)$ and
$B(r)$ $\to1$. On the other hand, the transfer of the body moving on
a circle in central symmetric homogeneous static gravitational
field from large distance to short distance according to
Kepler's  law means transformation from zero velocity to nonzero
velocity because of
$$\Delta t\to\Delta\tau=f(r)\Delta t,\quad
\Delta r\to\Delta R=\Delta r/f(r).$$

At last according to  geodesic equation
$$\frac{du^i}{ds}+\Gamma^i_{kl}u^ku^l=0,$$
where $\Gamma^i_{kl}$ -- Christoffel symbol in the metric
(\ref{Shwm}), from Newton equation we get
$$f^2(r)=1-\frac{r_g}r.$$

Thus the extention of field to the octonionic algebra and the
principles of STR give all the effects of GRG.

\section{Octonionic states}
The solution of the Dirac equation requires extension of bispinor to
the new algebra. For the sake of justice it should be noted that
this is a very big problem. The author reported the new solution of
this problem on the Conf. "RUSGRAV-13" \cite{DorConf}.

According to the principle of that extension the state is determinated
as a bispinor  $\Psi(x)$ from the set $BS\times O_4$:
\begin{equation}\label{Ps}
\Psi(x)=\left(\matrix{\Psi(x)_{in}^1\cr \Psi(x)_{in}^2\cr
\Psi(x)_{in}^3\cr \Psi(x)_{in}^4}\right),\qquad
\Psi(x)_{in}^{a'}=\left(\matrix{u(x)_{in}^{a'}&A(x)_{in}^{a'}\cr
B(x)_{in}^{a'}&v(x)_{in}^{a'}}\right),a'=1,2,3,4
\end{equation}
or
\begin{equation}\label{Ps2}
\Psi(x)=\left(\matrix{\left(\matrix{u(x)_{in}^1&A(x)_{in}^1\cr
B(x)_{in}^1&v(x)_{in}^1}\right)\cr
\left(\matrix{u(x)_{in}^2&A(x)_{in}^2\cr
B(x)_{in}^2&v(x)_{in}^2}\right)\cr
\left(\matrix{u(x)_{in}^3&A(x)_{in}^3\cr
B(x)_{in}^3&v(x)_{in}^3}\right)\cr
\left(\matrix{u(x)_{in}^4&A(x)_{in}^4\cr
B(x)_{in}^4&v(x)_{in}^4}\right)}\right)=
\left(\matrix{\left(\matrix{u(x)_{in}^{1,2}&A(x)_{in}^{1,2}\cr
B(x)_{in}^{1,2}&v(x)_{in}^{1,2}}\right)\cr
\left(\matrix{u(x)_{in}^{3,4}&A(x)_{in}^{3,4}\cr
B(x)_{in}^{3,4}&v(x)_{in}^{3,4}}\right)}\right)=
\left(\matrix{\varphi\cr\psi}\right).
\end{equation}

$v(x)^{a'},u(x)^{a'}$ -- scalar complex matrices $(2\times2)$ and
$A(x)^{a'}$, $B(x)^{a'}$ -- matrices: $B(x)^{a'}=\vec
b^{a'}(x)\cdot\vec\sigma,\vec b^{a'}(x)\in C$ (sign "$\cdot$" is
simple multiplication).

Considering $A=\vec a\vec\sigma,B=\vec b\vec\sigma$, we find the
probability density   $\rho=\Psi^+\Psi$ in the representation
$$\frac14tr\Psi^+\Psi=\frac14tr
\left(\matrix{\left(\matrix{u^*&-A\cr A^+&u}\right)
\left(\matrix{v^*&-B\cr B^+&v}\right)}\right)*
\left(\matrix{\left(\matrix{u&A\cr-A^+&u^*}\right)\cr
\left(\matrix{v&B\cr-B^+&v^*}\right)}\right)=$$
$$=|u|^2+|v|^2+|\vec a|^2+|\vec b|^2.$$

For example, we can consider the bispinor of electron in the form
\begin{equation}\label{Psel}
\Psi(x)=\left(\matrix{ \left(\matrix{u(x)^1&0\cr0&0}\right)\cr
\left(\matrix{u(x)^2&0\cr0&0}\right)\cr
\left(\matrix{u(x)^3&0\cr0&0}\right)\cr
\left(\matrix{u(x)^4&0\cr0&0}\right)}\right)=
\left(\matrix{u(x)^1\cr u(x)^2\cr u(x)^3\cr u(x)^4}\right).
\end{equation}
and the field
$A_\mu^{(0)}(x)=A_\mu(x),A_\mu^{(a)}(x)=0,a=1,2,3,\dots,7$.

The author believes that the weak and  strong interactions are
resulted as an extension of symmetry for $SU(2)$ and $SU(3)$ groups
\cite{Fradkin} where we can get $V-A$ interaction \cite{DorConf} but
the author wants to consider this problem in the next article.

\section{Free Lagrangian of this theory}
We define the free Lagrangian of field as
\begin{equation}\label{lagr}
L=-\frac14tr F_{ab}F^{ab},\qquad
F_{ab}=A_{b,a}-A_{a,b}-[A_aA_b-A_bA_a].
\end{equation}

Notice that  the Lagrangian doesn't equal to zero when its
associative part equals to zero because the nonassociative part
exists:
\begin{equation}\label{maincl}
L_{gr.}=-\frac12\eta^{bd}\eta^{ac}\varepsilon^{\tilde a\tilde
b\tilde c\tilde d}\left(A_a^{\tilde a}A_b^{\tilde
 b}-A_b^{\tilde a}A_a^{\tilde b}\right) \left(A_c^{\tilde c}A_d^{\tilde d}-
 A_d^{\tilde c}A_c^{\tilde d}\right).
\end{equation}

The last three multipliers in the formula (\ref{maincl}) are
similar to Riemannian tensor in the vierbein representation.
Particularly, the scalar curvature $R$ in the vierbein
representation is determined as
\begin{equation}\label{krtet}
R=\eta^{bd}\eta^{ac}R_{abcd}.
\end{equation}

The Ricci tensor $R_{abcd}$ has symmetry and antisymmetry properties
for indices as in (\ref{maincl}).

Thereby the analogy between nonassociative part of (\ref{lagr}) and
gravitation field Lagrangian is suggested. But it is necessary to
outline another property: Jacoby identity and cyclic rearrangement
by three indices don't take place. Therefore we don't come to
Einstein theory, but we have a complete coincidence between the
basic data experiments and our theory results.

In fact the equation on the octonionic algebra is equation for
differentials on a map. In this regard GRG offers to result general
equation in the whole atlas by the instrumentality of equation in
the Riemannian space.

\section{Acknowledgements}
The author expresses his gratitude to the participants of the
Friedmann Seminar for Theoretical Physics (St. Petersburg) for
profound discussions. The work was carried with the support of the
Russian Ministry of Education (grant RNP No 2.1.1.68.26).

\end{document}